\documentclass[lettersize,journal]{IEEEtran}
\usepackage{amsmath,amssymb,amsfonts}
\usepackage{algorithmic}
\usepackage{algorithm}
\usepackage{array}
\usepackage{subfig}
\usepackage{textcomp}
\usepackage{stfloats}
\usepackage{url}
\usepackage{verbatim}
\usepackage{graphicx}
\usepackage{cite}
\usepackage{color}
\usepackage{epstopdf}
\usepackage{makecell}
\DeclareMathAlphabet\mathbfcal{OMS}{cmsy}{b}{n}
\begin{document}

\title{Grid Evolution for Doubly Fractional Channel Estimation in OTFS Systems}

\author{Xiangjun Li, Pingzhi Fan, Qianli Wang, Zilong Liu
\thanks{Xiangjun Li, Pingzhi Fan and Qianli Wang are with the School of Info Sci \& Tech, Southwest Jiaotong University, Chengdu, China, and also with the Communications \& Sensor Networks for Modern Transportation (CSNMT) Int. Cooperation Research Centre of China, Chengdu, China. 
Zilong Liu is with the School of Computer Science and Electronics Engineering, University of Essex, U. K.
Corresponding author: Qianli Wang. Emails: lxj@my.swjtu.edu.cn; pzfan@swjtu.edu.cn; qianli\_wang@qq.com; zilong.liu@essex.ac.uk.}
}


\maketitle

\begin{abstract}
In orthogonal time-frequency space communications, the performances of existing on-grid and off-grid channel estimation (CE) schemes are determined by the delay-Doppler (DD) grid density. In practice, multiple real-life DD channel responses might be co-located within a same DD grid interval, leading to performance degradation. A finer grid interval is needed to distinguish these responses, but this could result in a significantly higher CE complexity when traditional methods are used.
To address this issue, a grid evolution method for doubly fractional CE is proposed by evolving the initially uniform coarse DD grid into a non-uniform dense grid. 
Simulation results show that our proposed method leads to improved computational efficiency, and achieves a good trade-off between CE performance and complexity.
\end{abstract}


\section{Introduction}

\IEEEPARstart{T}{he} next generation communication systems must support reliable information exchanges in highly dynamic environments, e.g., vehicle-to-everything systems, high-speed railways, etc. In these environments, one needs to deal with the fast time-varying channels incurred by high mobility. The  orthogonal frequency division multiplexing may be infeasible as its orthogonality can be easily destroyed by large Doppler. Against this background, orthogonal time-frequency space (OTFS) modulation has emerged in recent years due to its excellent performance in high mobility channels \cite{hadani1,P. Raviteja_1,P. Raviteja_2, Zhang, Y. Yang}. The basic principle of OTFS is to send the data symbols over delay-Doppler (DD) domain to achieve full-diversity. 


Channel estimation (CE) plays an important role in OTFS system design as it has direct impact to the detection performance. A number of OTFS CE schemes have been proposed.
An embedded frame structure and a threshold method are proposed in \cite{P. Raviteja_1} to estimate the channel.
In \cite{Shen,Zhao Lei}, the authors leveraged the sparsity in DD domain which permits the use of  orthogonal matching pursuit (OMP) and sparse Bayesian learning (SBL) for CE.
It is noted that these works assume that the real-life channel response is on the DD grid points. However, this assumption may not be valid in practice due to the fractional delay and Doppler values in real-world transmission. 
To address this problem, off-grid sparse Bayesian inference (OGSBI) \cite{Yang} was adopted in \cite{Wei}, leading to one-dimensional (1D) and two-dimensional (2D) off-grid CE scheme.
The 1D scheme demonstrates a good performance, but with high complexity. In contrast, the 2D scheme strikes a balance between complexity and performance. 
A 2D off-grid decomposition and SBL combination scheme was developed in \cite{Wang} to eliminate errors caused by interference between delay and Doppler. Yet it still suffers from a high complexity. 
These CE methods generally set a fixed initial grid. 
A denser initial grid usually achieves better CE performance, but it also brings in a significant computational burden.
Furthermore, multiple real-life channel responses may be co-located in the same DD grid interval even a denser grid is used and this could lead to deteriorated CE performance.

In this letter, inspired by the grid evolution (GE) method in \cite{Wang:GE}, a GE method for doubly fractional CE is proposed to tackle the above problems. 
Different from the method in \cite{Wang:GE}, our proposed GE is a 2D method, whereby the fission and adjustment of grid points are performed separately to avoid unnecessary calculations.
Compared with the GE scheme in \cite{Shan:GE}, the number of DD grid points in the proposed GE scheme is adjustable and hence it is generic and more flexible.
Specifically, the proposed GE scheme adaptively evolves from an initial uniform coarse grid to a non-uniform dense grid without a fixed interval.
The GE framework contains three processes, i.e., the learning, the fission and the adjustment. The learning process estimates the channel response at the grid points. The fission process adds new grid points along the delay or Doppler dimension, and separates multiple real-life channel responses into different DD grid intervals. 
The adjustment process combines the off-grid parameters into the current grid for decreasing the modeling error. These processes iterate alternately to achieve adaptive grid refinement and therefore obtain more accurate CE performance.
Compared with the previous schemes using a fixed uniform grid \cite{Zhao Lei,Wei,Wang}, the proposed GE scheme gives rise to significantly lower complexity due to smaller number of grid points.
And if similar number of grid points are taken, the proposed GE scheme shows better CE performance.

\section{System Models}
\subsection{Signal Model}
The DD channel response can be expressed as \cite{hadani1,P. Raviteja_1,P. Raviteja_2}
\begin{equation}
	h(\nu ,\tau )=\sum_{n=1}^P{h_n\delta \left( \nu -\nu _n \right) \delta \left( \tau -\tau _n \right)},
\end{equation}
where $P$ is the number of paths, $h_n$, $\tau _n\in \left( 0,\tau _{\max} \right) $ and $\nu _n\in \left( -\nu _{\max},\nu _{\max} \right) $ are the channel coefficient, delay, and Doppler of the $n$-th path, respectively. The symbol duration and bandwidth of OTFS system are $NT$ and $M\varDelta f$, respectively. $N$, $M$, $T$ and $\varDelta f$ are number of time slots, number of subcarriers, slots duration and subcarrier spacing, respectively.

In this letter, a single pilot $x_p$ with Doppler index $k_{p}^{\prime}$ and delay index $l_{p}^{\prime}$ in the DD grid is considered.
 To reduce interference between data and pilot, both the guard interval and the CE region in \cite{Wei} is used. The sizes of the CE region are $N_{\mathrm{T}}=\left( 2k_{\max}+1 \right) $ and $M_{\mathrm{T}}=\left( l_{\max}+1 \right) $ respectively, where $k_{\max}=\nu _{\max}NT$ and $l_{\max}=\tau _{\max}M\Delta f$. 
According to \cite{P. Raviteja_1,P. Raviteja_2,Wei}, the OTFS input-output relationship is
\begin{equation}\label{decoupled}
	y_{\mathrm{DD}}\left[ k,l \right] =x_p\sum_{n=1}^P{\tilde{h}_nw_{\nu}\left( k,k_{p}^{\prime},k_{\nu _n} \right) w_{\tau}^{\mathrm{H}}\left( l,l_{p}^{\prime},l_{\tau _n} \right)}+z\left[ k,l \right],
\end{equation}
where $y_{\mathrm{DD}}\left[ k,l \right] $ is the  received signals in the DD domain, $k\in \left\{ k_{p}^{\prime}-k_{\max},\cdots ,k_{p}^{\prime}+k_{\max} \right\} $, $l\in \left\{ l_{p}^{\prime},\cdots ,l_{p}^{\prime}+l_{\max} \right\} $, $z[k,l]\sim \mathcal{C} \mathcal{N} \left( 0,\lambda ^{-1} \right) $ is the noise and $\lambda ^{-1}$ is the variance, $\tilde{h}_n=h_ne^{-j2\pi \nu _n\tau _n}$, $k_{\nu _n}=\nu _nNT$, $l_{\tau _n}=\tau _nM\Delta f$, $w_{\nu}\left( \cdot \right) =w_{\tau}\left( \cdot \right) =\mathcal{F} \left( \cdot \right)$. In \cite{Wei,Wang}, $\mathcal{F} \left( \cdot \right)$ is denoted by
\begin{equation}
	\mathcal{F} \left( \eta ,\xi ,\gamma \right) =\frac{1}{Q}\left[ e^{-j(Q-1)\pi \frac{\eta -\xi -\gamma}{Q}}\frac{\sin \left( \pi \left( \eta -\xi -\gamma \right) \right)}{\sin \left( \frac{\pi \left( \eta -\xi -\gamma \right)}{Q} \right)} \right],
\end{equation}
where $Q=N$ for $w_{\nu}\left( \cdot \right) $ and $Q = M$ for $w_{\tau}\left( \cdot \right) $.

\subsection{Off-Grid Model and GE Model}
A finer grid is usually used for sparse representation as that of \cite{Wei}. In this case, the grid points of the initial uniform DD grid are $\left\{ \left\{ \bar{\boldsymbol{k}}_{\nu} \right\} \times \left\{ \bar{\boldsymbol{l}}_{\tau} \right\} \right\} \in \mathbb{R} ^{L_{\nu}\times L_{\tau}}$, $\bar{\boldsymbol{k}}_{\nu}=\left[ -k_{\max},\cdots ,k_{\max} \right] ^{\mathrm{T}}\in \mathbb{R} ^{L_{\nu}\times 1}$, $\bar{\boldsymbol{l}}_{\tau}=\left[ -l_{\max},\cdots ,l_{\max} \right] ^{\mathrm{T}}\in \mathbb{R} ^{L_{\tau}\times 1}$. Therefore, the initial delay and Doppler resolution are $r_{\nu}=\frac{2k_{\max}}{L_{\nu}-1}$ and $r_{\tau}=\frac{l_{\max}}{L_{\tau}-1}$, respectively.
Let the number of grid points in the sampled DD grid be $L=L_{\nu}L_{\tau}$. After vectorizing the DD grid, all DD grid points can be expressed as $\tilde{\boldsymbol{S}}=\left\{ \tilde{\boldsymbol{k}}_{\nu},\tilde{\boldsymbol{l}}_{\tau} \right\} $, $\tilde{\boldsymbol{l}}_{\tau}=\left[ l_0,l_1,\cdots l_i,\cdots ,l_{L-1} \right] ^{\mathrm{T}}$, $\tilde{\boldsymbol{k}}_{\nu}=\left[ k_0,k_1,\cdots ,k_i,\cdots ,k_{L-1} \right] ^{\mathrm{T}}$. Let $i\in \left\{ 1,\cdots L \right\}$, $\boldsymbol{w}_{\nu}\left( k_i \right) =\left[ w_{\nu}\left( k_{p}^{\prime}-k_{\max},k_{p}^{\prime},k_i \right) ,\cdots ,w_{\nu}\left( k_{p}^{\prime}+k_{\max},k_{p}^{\prime},k_i \right) \right] ^{\mathrm{T}}\in \mathbb{C} ^{N_{\mathrm{T}}\times 1}$, $\boldsymbol{w}_{\tau}\left( l_i \right) =\left[ w_{\nu}\left( l_{p}^{\prime},l_{p}^{\prime},l_i \right) ,\cdots ,w_{\nu}\left( l_{p}^{\prime}+l_{\max},l_{p}^{\prime},l_i \right) \right] ^{\mathrm{T}}\\\in \mathbb{C} ^{M_{\mathrm{T}}\times 1}$. Then, the on-grid model based on Eq. \eqref{decoupled} is 
\begin{equation}\label{conventional observation model}
	\boldsymbol{y}=x_p\boldsymbol{\varPhi }_{\mathrm{I}}\left( \tilde{\boldsymbol{S}} \right) \tilde{\boldsymbol{h}}+\boldsymbol{z},
\end{equation}
where $\boldsymbol{y},\boldsymbol{z}\in \mathbb{C} ^{N_{\mathrm{T}}M_{\mathrm{T}}\times 1}$, $\boldsymbol{y}=\mathrm{vec}\left( \boldsymbol{Y} \right) $, $\boldsymbol{Y}\in \mathbb{C} ^{N_T\times M_T}$ is the matrix form of $y_{DD}\left[ k,l \right] $, $\tilde{\boldsymbol{h}}\in \mathbb{C} ^{L\times 1}$, 
$\boldsymbol{\varPhi }_{\mathrm{I}}\left( \tilde{\boldsymbol{S}} \right) =\boldsymbol{\psi }\odot \left[ \boldsymbol{\phi }_{\mathrm{I}}\left( k_0,l_0 \right) ,\cdots ,\boldsymbol{\phi }_{\mathrm{I}}\left( k_{L-1},l_{L-1} \right) \right] \in \mathbb{C} ^{N_{\mathrm{T}}M_{\mathrm{T}}\times L}$ is the on-grid part of the measurement matrix corresponding to the current DD grid, $\boldsymbol{\phi }_{\mathrm{I}}\left( k_i,l_i \right) =\mathrm{vec}\left( \boldsymbol{w}_{\nu}\left( k_i \right) \boldsymbol{w}_{\tau}^{H}\left( l_i \right) \right) \in \mathbb{C} ^{N_{\mathrm{T}}M_{\mathrm{T}}\times 1}$, 
$\boldsymbol{\psi }=\left[ \boldsymbol{\varphi }_0,\cdots ,\boldsymbol{\varphi }_{L-1} \right] \in \mathbb{C} ^{N_{\mathrm{T}}M_{\mathrm{T}}\times L}$, 
$\boldsymbol{\varphi }_i=e^{\frac{-j2\pi k_il_i}{NM}}\mathbf{1}$, 
$\mathbf{1}\in \mathbb{C} ^{N_{\mathrm{T}}M_{\mathrm{T}}\times 1}$ is the all-ones column vector, 
$\mathrm{vec}\left( \cdot \right)$ is the vectorization operation, $\odot $ is the dot product. 

Since in practice the channel response generally does not fall exactly on the sampled DD grid points, an off-grid CE scheme was proposed in \cite{Wei,Wang}. 
Denote by $\left( k_i\in \tilde{\boldsymbol{k}}_{\nu},l_i\in \tilde{\boldsymbol{l}}_{\tau} \right) $ the $i$-th grid point which is closest to $(k_{\nu _n},l_{\tau _n})$ in the discrete DD plane. Then a linear approximation in \cite{Wei} can be obtained by first-order Taylor expansion, i.e,
\begin{equation}\label{Taylor}
	\begin{split}
		\boldsymbol{\varPhi }\left( k_{\nu _n},l_{\tau _n} \right) =\boldsymbol{\phi }_{\mathrm{I}}\left( k_i,l_i \right) +\boldsymbol{\phi }_{\nu}\left( k_i,l_i \right) \kappa _i+\boldsymbol{\phi }_{\tau}\left( k_i,l_i \right) \iota _i\\+o\left( \kappa _i \right) +o\left( \iota _i \right),
	\end{split}
\end{equation}
where $\kappa _i=k_{\nu _n}-k_i$, $\iota _i=l_{\tau _n}-l_i$, $o\left( \kappa _i \right) $ and $o\left( \iota _i \right) $ are the modeling errors, 
$\boldsymbol{\phi }_{\nu}\left( k_i,l_i \right) =\mathrm{vec}\left( \boldsymbol{w}_{\nu}^{\prime}\left( k_i \right) \boldsymbol{w}_{\tau}^{\mathrm{H}}\left( l_i \right) \right)$, $\boldsymbol{\phi }_{\tau}\left( k_i,l_i \right) =\mathrm{vec}\left( \boldsymbol{w}_{\nu}\left( k_i \right) \left( \boldsymbol{w}_{\tau}^{\prime}\left( l_i \right) \right) ^{\mathrm{H}} \right)$, $\boldsymbol{w}_{\nu}^{\prime}\left( k_i \right)$ and $\boldsymbol{w}_{\tau}^{\prime}\left( l_i \right)$ are the partial derivatives with respect to $k_i$ and $l_i$ respectively. Here, $\kappa _i$ and $\iota _i$ are assumed to follow uniform distributions
\begin{equation}
	\kappa _i\sim \mathcal{U} \left[ -\frac{1}{2}\tilde{r}_{\nu _-},\frac{1}{2}\tilde{r}_{\nu _+} \right] ,  \iota _i\sim \mathcal{U} \left[ -\frac{1}{2}\tilde{r}_{\tau _-},\frac{1}{2}\tilde{r}_{\tau _+} \right],
\end{equation}
where $\tilde{r}_{\nu}$ and $\tilde{r}_{\tau}$ respectively represent the Doppler and delay resolutions between a grid point and its adjacent grid points, $+$ and $-$ denote positive and negative directions respectively.

Let $\tilde{\boldsymbol{\kappa }}=\left[ \kappa _0,\kappa _1,\cdots ,\kappa _{L-1} \right] ^{\mathrm{T}}\in \mathbb{C} ^{L\times 1}$, 
$\tilde{\boldsymbol{\iota }}=\left[ \iota _0,\iota _1,\cdots ,\iota _{L-1} \right] ^{\mathrm{T}}\in \mathbb{C} ^{L\times 1}$. Similar to $\boldsymbol{\varPhi }_{\mathrm{I}}\left( \tilde{\boldsymbol{S}} \right) $, $\boldsymbol{\phi }_{\nu}\left( k_i,l_i \right)$ and $\boldsymbol{\phi }_{\tau}\left( k_i,l_i \right)$ can be arranged in order to obtain $\boldsymbol{\varPhi }_{\nu}\left( \tilde{\boldsymbol{S}} \right) $ and $\boldsymbol{\varPhi }_{\tau}\left( \tilde{\boldsymbol{S}} \right) $, respectively.
Then the measurement matrix $\boldsymbol{\varPhi }\left( \tilde{\boldsymbol{S}},\tilde{\boldsymbol{\kappa}},\tilde{\boldsymbol{\iota}} \right) \in \mathbb{C} ^{N_{\mathrm{T}}M_{\mathrm{T}}\times L}$ can be expressed as
\begin{equation}
	\boldsymbol{\varPhi }\left( \tilde{\boldsymbol{S}},\tilde{\boldsymbol{\kappa}},\tilde{\boldsymbol{\iota}} \right) =\boldsymbol{\varPhi }_{\mathrm{I}}\left( \tilde{\boldsymbol{S}} \right) +\boldsymbol{\varPhi }_{\nu}\left( \tilde{\boldsymbol{S}} \right) \mathrm{diag}\left\{ \tilde{\boldsymbol{\kappa}} \right\} +\boldsymbol{\varPhi }_{\tau}\left( \tilde{\boldsymbol{S}} \right) \mathrm{diag}\left\{ \tilde{\boldsymbol{\iota}} \right\},
\end{equation}
where $\mathrm{diag}\left\{ \cdot \right\}$ is the diagonal matrix operator. After absorbing the approximation error into the noise, the observation model can be written as
\begin{equation}\label{off-grid model}
	\boldsymbol{y}=x_p\boldsymbol{\varPhi }\left( \tilde{\boldsymbol{S}},\tilde{\boldsymbol{\kappa}},\tilde{\boldsymbol{\iota}} \right) \tilde{\boldsymbol{h}}+\boldsymbol{z}.
\end{equation}

Note that $\tilde{\boldsymbol{S}}$ is constant in both the on-grid and off-grid models, meaning that $r_{\nu}$ and $r_{\tau}$ are constants. As mentioned before, the performance of the on-grid and off-grid models will deteriorate when multiple real-life channel responses are located in the same DD interval. Therefore, the proposed GE scheme uses a varying $\boldsymbol{S}$ to decrease the modeling error, where $\boldsymbol{S}=\left\{ \tilde{\boldsymbol{S}}_{\mathrm{ini}},\tilde{\boldsymbol{S}}_{\mathrm{GE}} \right\}$, the $\tilde{\boldsymbol{S}}_{\mathrm{ini}}$ is the $\boldsymbol{S}$ with $r_{\nu}=r_{\tau}=1$, $\tilde{\boldsymbol{S}}_{\mathrm{GE}}$ represents the added grid points in the GE scheme. Therefore, the model in GE is,
\begin{equation}\label{GE model}
	\boldsymbol{y}=x_p\boldsymbol{\varPhi }\left( \boldsymbol{S},\boldsymbol{\kappa },\boldsymbol{\iota } \right) \boldsymbol{h}+\boldsymbol{z},
\end{equation}
where the sizes of $\boldsymbol{\kappa}$, $\boldsymbol{\iota}$ and $\boldsymbol{h}$ are corresponding to $\boldsymbol{S}$.

\section{Grid Evolution Channel Estimation}
\begin{figure}[t]
	\centerline{\includegraphics[width = 7cm ]{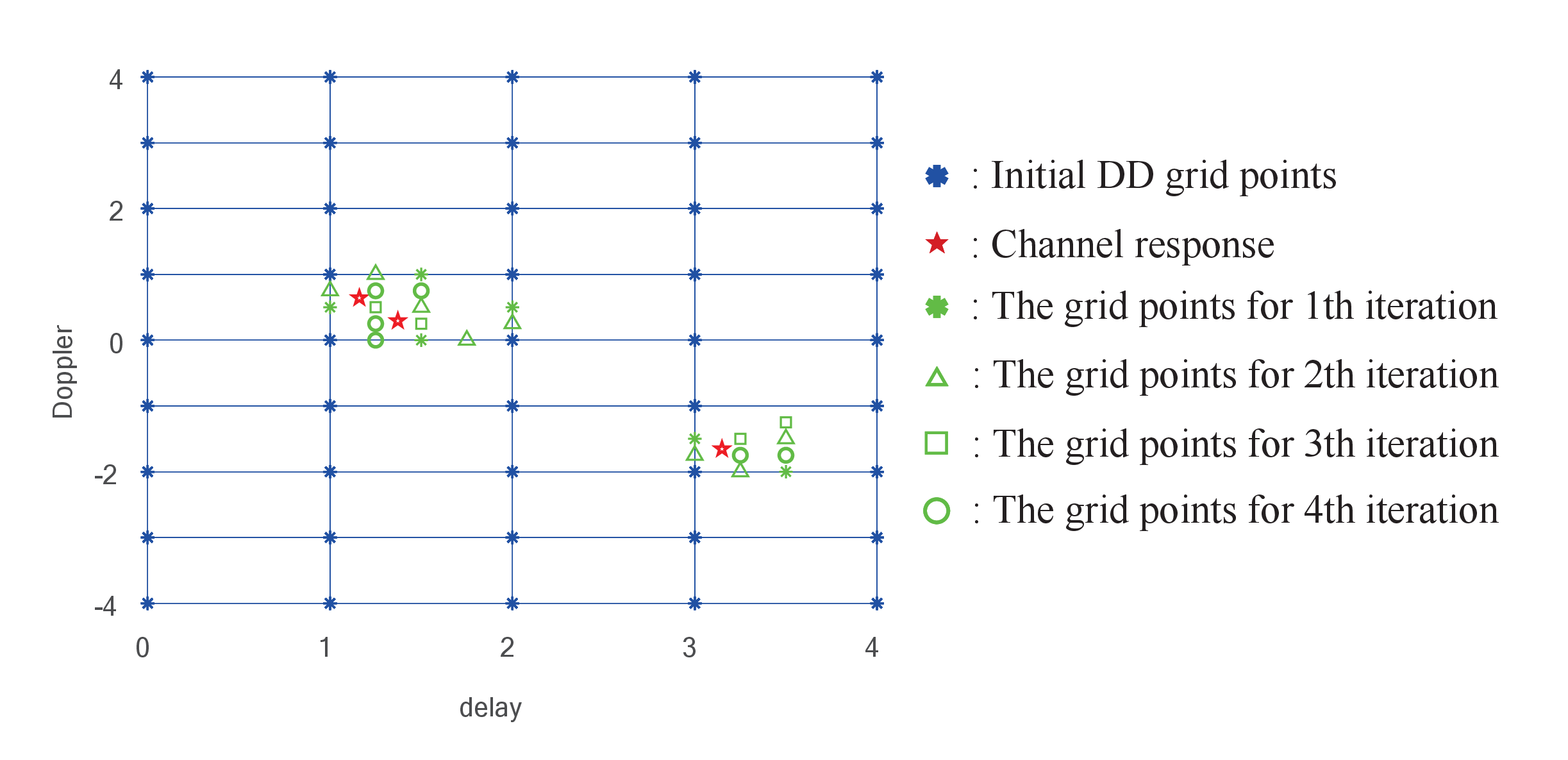}}
	\caption{The fission process for the GE scheme.}
	\label{GE-fission}
\end{figure}

To get the grid points $\boldsymbol{S}$ and estimate $\boldsymbol{h}$ in Eq. \eqref{GE model}, GE method is adopted to increase the local resolution for a more accurate representation of the sparse channel responses in the DD domain. As shown in Fig. \ref{GE-fission}, for the off-grid scheme using a coarse DD grid, the modeling error will be large. There are two sources of modeling error. One is that larger off-grid gaps lead to large $o\left( \kappa _i \right) $ and $o\left( \iota _i \right) $. The second is because multiple channel responses are located in the same DD interval and they may not be distinguished. By using the GE method, more grid points will be generated around the true responses by decreasing the off-grid gaps to distinguish the close paths. Note that this procedure is different from direct grid refinement like that of chirp z-transform. The GE method is based on the compressive sensing thus the coefficients corresponding to these grid points will have to meet the sparsity constraint, resulting a sparse representation of the channel responses.

\begin{figure}[t]
	\centerline{\includegraphics[width = 7cm ]{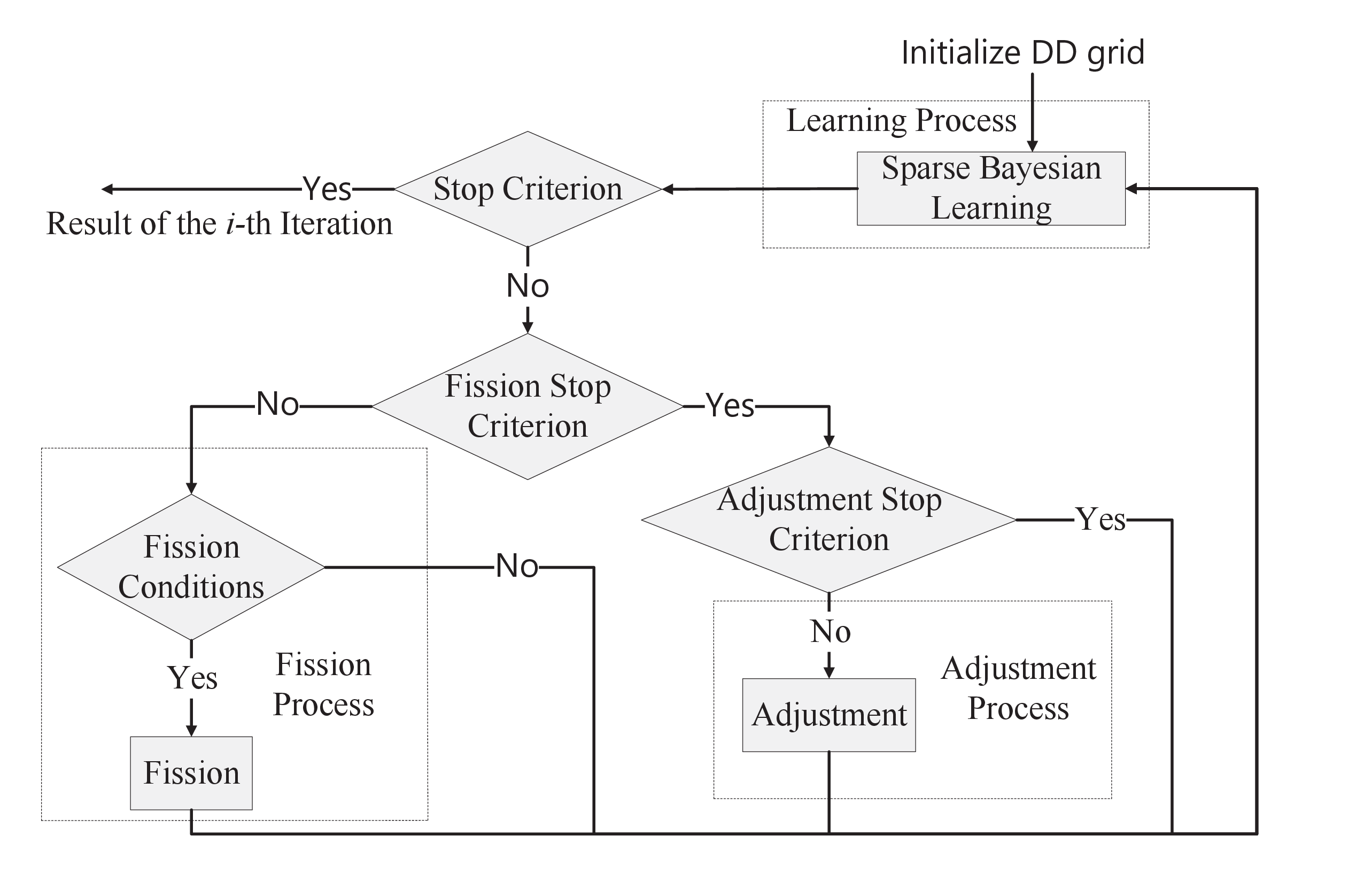}}
	\caption{The procedure of GE channel estimation.}
	\label{GE}
\end{figure}

The GE method consists of the learning, the fission and the adjustment processes, as shown in Fig. \ref{GE}. The fission process can distinguish multiple channel responses within the same initial DD interval. The adjustment process combines the off-grid parameters into the current grid to reduce the off-grid gap. The learning process estimates the channel response at the grid points by SBL. Alternate iterations of the learning-fission and alternate iterations of the learning-adjustment are performed sequentially to adaptively refine the DD grid for improving the local DD resolution and reducing modeling error.
To strike a trade-off between modeling error and computational workload, stop criteria are proposed for fission and adjustment.

\subsection{Learning Process}
OGSBI in \cite{Yang} is used to estimate the channel parameters in the learning process. 
According to \cite{Yang,Wei}, the posterior conditional distribution of $\boldsymbol{h}$ is assumed to be
\begin{equation}
	p\left( \boldsymbol{h}\left| \boldsymbol{y};\boldsymbol{\alpha },\boldsymbol{\kappa },\boldsymbol{\iota },\lambda \right. \right) =\mathcal{C} \mathcal{N} \left( \boldsymbol{h}\left| \boldsymbol{\mu },\boldsymbol{\varSigma } \right. \right),
\end{equation}
where $\boldsymbol{\alpha }$ is the hyperparameter that models the sparsity of $\boldsymbol{h}$ and follows a Gamma distribution controlled by the parameter $\rho $,
$\lambda $ follows a Gamma distribution determined by $a$ and $b$,
The mean $\boldsymbol{\mu}$ and covariance $\boldsymbol{\varSigma }$ for the $(k)$-th iteration are 
\begin{equation}\label{u}
	\boldsymbol{\mu }^{\left( k \right)}=\lambda ^{\left( k \right)}\boldsymbol{\varSigma }^{\left( k \right)}\boldsymbol{\varPhi }^{\mathrm{H}}\left( \boldsymbol{S}^{\left( k \right)},\boldsymbol{\kappa }^{\left( k \right)},\boldsymbol{\iota }^{\left( k \right)} \right) \boldsymbol{y},
\end{equation}
\begin{equation}\label{Sigma}
	\begin{aligned}
		\boldsymbol{\varSigma }^{\left( k \right)}=\mathrm{diag}\left( \boldsymbol{\alpha }^{\left( k \right)} \right) -\left( \boldsymbol{\alpha }^{\left( k \right)}\odot \left( \boldsymbol{\alpha }^{\left( k \right)}\odot \boldsymbol{C} \right) ^{\mathrm{T}} \right) ^{\mathrm{T}},
	\end{aligned}
\end{equation}
where $\boldsymbol{C}=\boldsymbol{\varPhi }^{\mathrm{H}}\left( \boldsymbol{S}^{\left( k \right)},\boldsymbol{\kappa }^{\left( k \right)},\boldsymbol{\iota }^{\left( k \right)} \right) \boldsymbol{\varSigma }_{\mathrm{y}}^{-1}\boldsymbol{\varPhi }\left( \boldsymbol{S}^{\left( k \right)},\boldsymbol{\kappa }^{\left( k \right)},\boldsymbol{\iota }^{\left( k \right)} \right) $,
$\boldsymbol{\varSigma }_{\mathrm{y}}=\boldsymbol{\varPhi }\left( \boldsymbol{S}^{\left( k \right)},\boldsymbol{\kappa }^{\left( k \right)},\boldsymbol{\iota }^{\left( k \right)} \right) \left( \boldsymbol{\alpha }^{\left( k \right)}\circledcirc \boldsymbol{\varPhi }^{\mathrm{H}}\left( \boldsymbol{S}^{\left( k \right)},\boldsymbol{\kappa }^{\left( k \right)},\boldsymbol{\iota }^{\left( k \right)} \right) \right) +\lambda ^{\left( k \right)}\boldsymbol{I}$,
$\boldsymbol{\alpha }\circledcirc \boldsymbol{C}$ is the dot product of the elements of the column vector $\boldsymbol{\alpha }$ with the corresponding rows of $\boldsymbol{C}$. Different from the traditional update formulation of $\boldsymbol{\varSigma }$, the diagonal matrix property is considered here to reduce the complexity.

The expectation–maximization algorithm is used to update each hyper-parameter for the $(k+1)$-th iteration, i.e.,
\begin{equation}\label{alpha}
	\alpha \left( i \right) ^{\left( k+1 \right)}=\frac{\sqrt{1+4\rho \left( \varSigma \left( i,i \right) ^{\left( k \right)}+\left| \mu \left( i \right) ^{\left( k \right)} \right|^2 \right)}-1}{2\rho},
\end{equation}
\begin{equation}
	\lambda ^{\left( k+1 \right)}=\frac{2a-2+M_{\mathrm{T}}N_{\mathrm{T}}}{2b+\varDelta {y}},
\end{equation}
where $i\in \left\{ 1,\cdots ,L \right\}$, $\varDelta y=\left( \lambda ^{\left( k \right)} \right) ^{-1}\sum_{i=1}^L{1-\small{\frac{\varSigma \left( i,i \right) ^{\left( k \right)}}{\alpha \left( i \right) ^{\left( k \right)}}}}+\left\| \boldsymbol{y}-\boldsymbol{\varPhi }\left( \boldsymbol{S}^{\left( k \right)},\boldsymbol{\kappa }^{\left( k \right)},\boldsymbol{\iota }^{\left( k \right)} \right) \boldsymbol{\mu }^{\left( k \right)} \right\| ^2$, $\lambda>0$, $a>0$, $b>0$. For updates of off-grid parameters, please refer to \cite{Yang,Wei}.
The iteration will stop if $\frac{\left\| \boldsymbol{\alpha }^{\left( k+1 \right)}-\boldsymbol{\alpha }^{\left( k \right)} \right\|}{\left\| \boldsymbol{\alpha }^{\left( k \right)} \right\|}<\delta $ or the maximum number of iterations $K$ is reached, where $\delta $ is a tolerance.

\subsection{Fission Process}
Different from the 1D angular grid in \cite{Wang:GE}, we consider 2D grid in this paper. If the fission is carried as that of \cite{Wang:GE}, the grid $\boldsymbol{S}$ will become $\left\{ \left\{ \bar{\boldsymbol{k}}_{\nu},\boldsymbol{k}_{\mathrm{new}} \right\} \times \left\{ \bar{\boldsymbol{l}}_{\tau},\boldsymbol{l}_{\mathrm{new}} \right\} \right\}$, which is equivalent to inserting several rows and several columns of grid points into the initial DD grid.
Obviously, except for a few points which are close to the channel responses, many grid points may be irrelevant to the channel responses, thus leading to extra calculation cost.

As shown in Fig. \ref{GE-fission}, we add grid points around the true channel responses. 
Three fission conditions are used like that of \cite{Wang:GE} to decide whether a fission of a grid should happen. For all the grid points $i\in \left\{ 1,\cdots ,L \right\}$, grid point with $\left| \mu \left( i \right) \right|>\varepsilon \sqrt{\lambda ^{-1}}$ needs fission for a higher representation accuracy, where $\varepsilon$ is a weight related to miss detection or false-alarm probabilities. Only the grid point corresponding to the local maximum value can be selected for fission to prevent too many fission grid points corresponding to a large channel response. Finally, $\tilde{r}_{\nu}>r_{\min}$ or $\tilde{r}_{\tau}>r_{\min}$ should be satisfied.

When these three conditions are met, the fission needs to be carried out simultaneously in both the delay and the Doppler dimensions based on estimated off-grid parameters.
Assume that the chosen grid point $\chi$ is $\left( k_i,l_i \right) $, and the estimation result of corresponding off-grid parameter is $\left( \hat{\kappa}_i,\hat{\iota}_i \right) $. The off-grid parameters can indicate that the channel response is in a certain direction of grid point $\chi$. Thus two new grid points $\chi _{k}=\left( k_{\mathrm{fission}},l_i \right)$ and $\chi _{l}=\left( k_i,l_{\mathrm{fission}} \right)$ will be generated after fission, where $k_{\mathrm{fission}}$ and $l_{\mathrm{fission}}$ are
\begin{equation}\label{k-fission}
	k_{\mathrm{fission}}=\left\{ \begin{array}{c}	k_i+\small{\frac{1}{2}}\tilde{r}_{\nu _+},\hat{\kappa}_i>0\\	k_i-\small{\frac{1}{2}}\tilde{r}_{\nu _-},\hat{\kappa}_i<0\\\end{array} \right. ,\\
\end{equation}
\begin{equation}\label{l-fission}
	l_{\mathrm{fission}}=\left\{ \begin{array}{c}	l_i+\small{\frac{1}{2}}\tilde{r}_{\tau _+},\hat{\iota}_i>0\\	l_i-\small{\frac{1}{2}}\tilde{r}_{\tau _-},\hat{\iota}_i<0\\\end{array} \right. .\\
\end{equation}

If a grid point fissions, it will be necessary to add 1 or 2 elements  corresponding to the grid point for $\boldsymbol{\kappa }$, $\boldsymbol{\iota }$ and $\boldsymbol{h}$ in Eq. \eqref{GE model}. 
Owing to the OGSBI method we used for estimation, the fission of prior of $\boldsymbol{h}$, i.e., $\alpha$, is needed.
Assume that the $\alpha$ value corresponding to grid points $\chi$, $\chi _{\mathrm{f}}$, $\chi _{k}$ and $\chi _{l}$ are $\alpha _{\chi}$, $\alpha _{\chi _{\mathrm{f}}}$, $\alpha _{\chi _k}$ and $\alpha _{\chi _l}$, respectively, where $\chi _{\mathrm{f}}$ is the grid point $\chi$ version after fission. To keep the prior $p\left( \boldsymbol{\alpha } \right) $ unchanged after fission, $p\left( \boldsymbol{\alpha } \right) $ should be a constant \cite{Wang:GE}. Therefore,
\begin{equation}
	\alpha _{\chi_\mathrm{f}}=\alpha _{\chi _k}=\alpha _{\chi _l}=\frac{1}{3}\alpha _{\chi},
\end{equation}
where $\alpha _{\chi}$ is estimated by Eq. \eqref{alpha}. 

Note that too many fissions may reduce computational efficiency or even deteriorate the CE performance because of the high correlation between the grid points. Therefore, beyond the fission conditions, the fission stop criterion is introduced that no more fission will happen if the maximum number of iterations $K_{\mathrm{f}}$ for fission process is reached. $K_{\mathrm{f}}$ may be larger than $\log _2(1/r_{\min})$, which refers to the average iteration number for reaching a satisfying resolution.

\subsection{Adjustment Process}
If a dense grid is used in the uniform grid model, the Taylor approximation errors would be small. For the GE scheme, after the fission process, a non-uniform and locally fine DD grid with better CE performance will be obtained. However, the off-grid gap of the current DD grid cannot be ignored. This is because the CE performance may be affected by the fractional channel. Therefore, Adjustment process is proposed in this letter to further reduce the approximation errors. 


It is not necessary that all grid points are adjusted.
For all the grid points $i\in \left\{ 1,\cdots ,L \right\} $, if $\left| \mu \left( i \right) \right|>\varepsilon \sqrt{\lambda ^{-1}}$, then we adjust the grid. 
The adjustment of a DD grid is given by
\begin{equation}\label{correction}
	\left\{ \begin{array}{l}	k_i\gets k_i+\kappa _i\\	l_i\gets l_i+\iota _i\\\end{array} \right. .
\end{equation}
When the adjustment process is completed, the off-grid parameters $k_i$ and $l_i$ will be reduced, leading to decreased Taylor approximation errors $o\left( \kappa _i \right) $ and $o\left( \iota _i \right) $. Since $\kappa _i$ and $\iota _i$ are integrated into the delay and Doppler values of a grid point, $\boldsymbol{\varPhi }\left( \boldsymbol{S},\boldsymbol{\kappa },\boldsymbol{\iota } \right)$ should be recalculated but this incurs a higher complexity. In this case, an adjustment stop criterion is required. The iteration of the adjustment process will stop if $\left\| \left[ \boldsymbol{\kappa }^{\left( k \right)},\boldsymbol{\iota }^{\left( k \right)} \right] \right\| <\delta _{\mathrm{a}}$ or the maximum number of iterations $K_{\mathrm{a}}$ for the adjustment process is reached, where $\delta _{\mathrm{a}}$ is a tolerance.

In addition to the above framework, another framework is that fission and adjustment are performed in the same iteration or execute iteratively. 
This letter does not use these two frameworks because the initial grid in GE scheme is coarse and it is quite possible that multiple paths are in a same grid interval.
Therefore, a gird point in early iterations may represent a response of multiple paths and undergo fission in future iterations.
After fission, the grid distribution will be reset thus the computational workload of adjustment of these grid point is wasted.    
In our framework, the paths are assumed to be distinguishable and represented by different grid points after the stop of fission process. Then the adjustment process could decrease the Taylor approximation errors for each path.

%
%
%

\section{Complexity Analysis and Simulation Results}
The proposed GE scheme is compared with the on-grid, 1D off-grid \cite{Wei}, 2D off-grid-U \cite{Wei} and 2D off-grid \cite{Wang}.
Table \ref{parameters} presents the simulation parameters. The pilot pattern and guard interval in \cite{Wei} are used. The power of single pilot is $30$ dB higher than the average power of data. In GE, $a=b=10^{-4}$, $\rho =10^{-2}$, $\delta ={{10}^{-3}}$, $\delta _{\mathrm{a}}=10^{-1}$, $K_{\mathrm{f}}=5$, $K_{\mathrm{a}}=50$, $K=200$. If not stated otherwise, the resolution of the uniform grid scheme is $r=r_{\nu}=r_{\tau}=\frac{1}{4}$. The initial resolution in the GE scheme is $1$, and $r_{\min}=\frac{1}{4}$. 

\begin{table}[t]\caption{Simulation parameters}
	\centering
	\label{parameters}
	\begin{tabular}{|c|c|}
		\hline
		Parameter   & Value   \\\hline
		DD grid size& $N=M=32$\\\hline
		Carrier frequency& $fc\text{ }=\text{ }4\text{ GHz}$\\\hline
		Subcarrier spacing & $\Delta f\text{ }=\text{ }15\text{ kHz}$\\\hline
		Number of channel paths   & $P\text{ }=\text{ }5$ \\\hline
		The channel coefficient  & $h_i\sim \mathcal{C} \mathcal{N} \left( 0,\frac{\exp \left( -0.1l_{\tau _i} \right)}{\sum\nolimits_i^{}{\exp \left( -0.1l_{\tau _i} \right)}} \right) $ \\\hline
		Maximum delay & ${{\tau }_{\max }}=8.3\times {{10}^{-6}}\text{ s}$ \\\hline
		Maximum relative velocity   &  $500\text{ km/h}$\\\hline
	\end{tabular}
\end{table}

\begin{figure}[h]
	\centerline{\includegraphics[width = 7cm ]{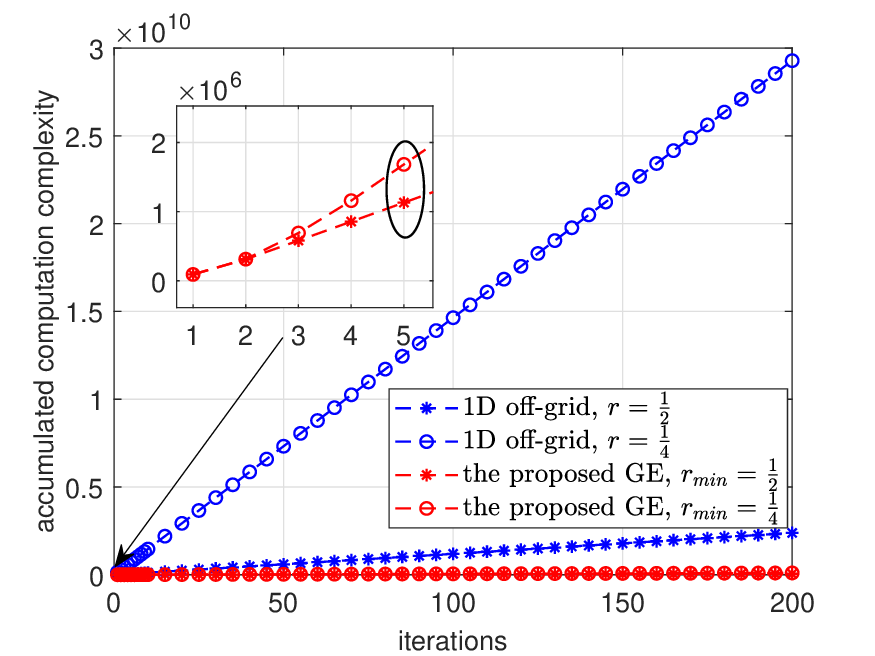}}
	\caption{Accumulated computational complexity comparison.}
	\label{computation complexity}
\end{figure}
The accumulated complexity comparison at different numbers of iterations is shown in Fig. \ref{computation complexity}.
The main complexity of the GE scheme is the same as that of the scheme in \cite{Wei}. For the GE scheme, the initial uniform DD grid is coarse and $L$ is small. The fission process stops at the iteration number in the black circle and the final $L$ is still small. This leads to significantly reduced complexity of the proposed GE scheme.

\begin{figure}[h]
	\centerline{\includegraphics[width = 7cm ]{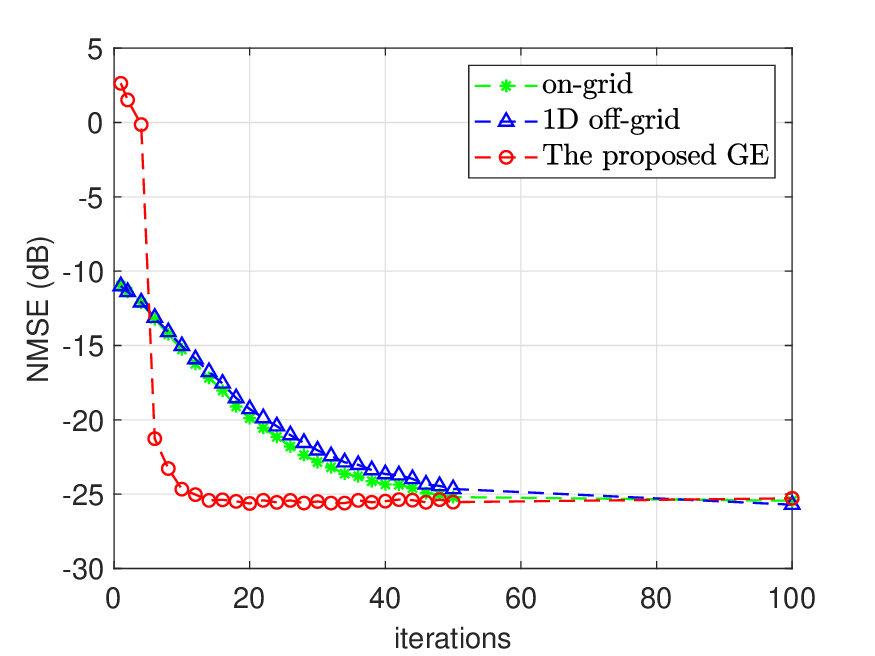}}
	\caption{Convergence comparison between the GE scheme and the uniform grid scheme.}
	\label{NMSE-iters}
\end{figure}

In Fig. \ref{NMSE-iters}, the convergence of the different schemes is compared. Compared with the other two uniform grid schemes, the GE scheme can converge faster. Combining Fig. \ref{computation complexity} and \ref{NMSE-iters}, it can be seen that with the same number of iterations, GE scheme has lower computational complexity and faster convergence while ensuring the NMSE performance.

\begin{figure}[t]
	\centerline{\includegraphics[width = 7cm ]{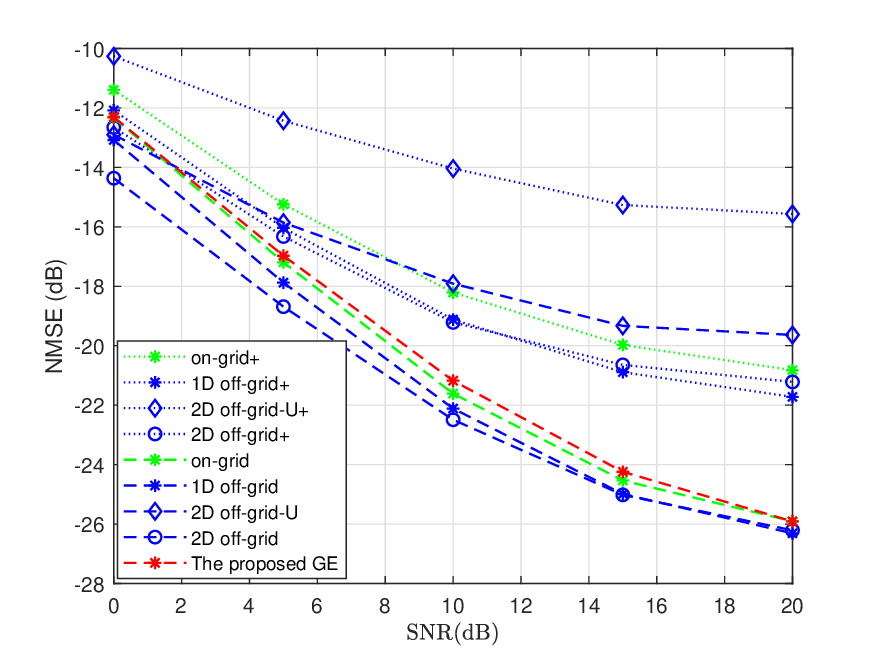}}
	\caption{NMSE comparison}
	\label{NMSE SNR}
\end{figure}

The NMSE comparison between different methods is given in Fig. \ref{NMSE SNR}.
The expected resolution is $r=\frac{1}{4}$, thus the number of grid points for all the uniform grid schemes is 561. The initial number of grid points for the GE scheme is 45 and finally reaches around 110. 
For a fair comparison, uniform grid schemes marked with $'+'$ are also presented and this indicates that 
the number of grid points, in these uniform grid schemes, is approximately equal to 110 as that in the GE scheme. 
It can be seen that the performance of the proposed GE with much fewer grid points can be very close to that of the uniform grid scheme with $r=\frac{1}{4}$, and significantly outperforms the uniform grid schemes when the number of grid points is comparable. 

\begin{figure}[h]
	\centerline{\includegraphics[width = 7cm ]{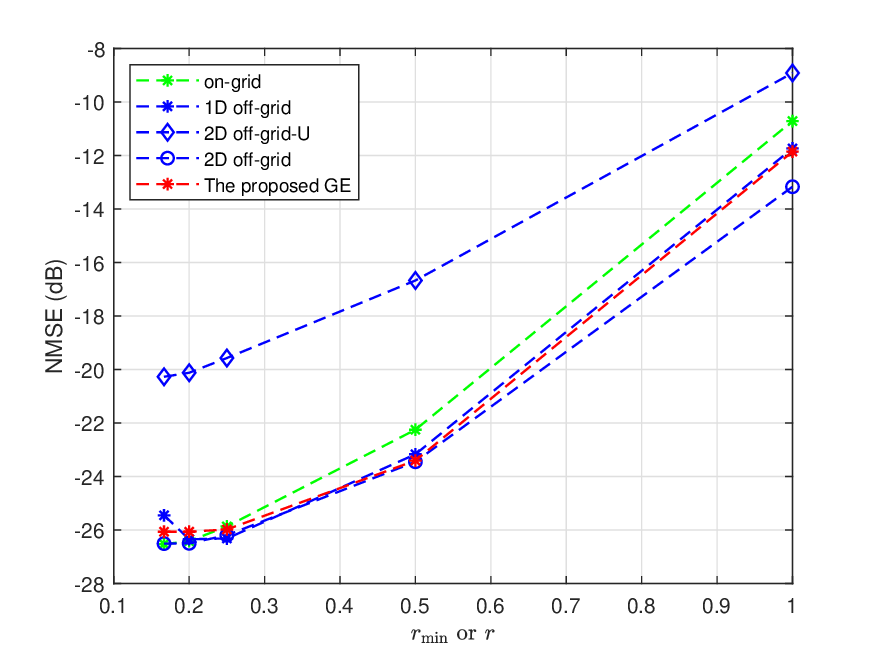}}
	\caption{NMSE under different $r_{\min}$ or $r$.}
	\label{NMSE-r_min}
\end{figure}

Note that the performance of GE method relies on the choice of $r_{\min}$. Therefore, to demonstrate the influence of $r_{\min}$, Fig. \ref{NMSE-r_min} shows the comparison of NMSE for different $r_{\min}\,\,\mathrm{and}\,\,r$ at $\mathrm{SNR}=20$ dB.
It is found that the NMSE of GE is smaller if we take a smaller $r_{\min}$. And except for the 2D off-grid scheme that considers the tandem off-grid distortion to make the model error smaller, the GE scheme has the best performance when $r$ is large. Considering all simulation results together, the GE scheme achieves a good trade-off between CE performance and complexity.

\section{Conclusions}
An improved GE method for doubly fractional CE has been proposed in this letter. SBL and grid refinement are combined to adaptively evolve uniform DD grid into non-uniform grid. The key idea of the proposed GE method consists of three processes, i.e., learning, fission and adjustment process. The learning process estimates the channel response at the current grid points, the fission process adds new grid points, and the adjustment process employs off-grid parameters to adjust the DD grid. The learning process iterates alternately with the other two processes, respectively. Simulation results show that the proposed GE scheme achieves an excellent tradeoff between the CE performance and complexity.

\newpage

 
\vspace{11pt}
%
%
%

\vfill

\end{document}